\documentclass[sn-mathphys,Numbered]{sn-jnl}

\usepackage{graphicx}%
\usepackage{multirow}%
\usepackage{amsmath,amssymb,amsfonts}%
\usepackage{amsthm}%
\usepackage{mathrsfs}%
\usepackage[title]{appendix}%
\usepackage{xcolor}%
\usepackage{textcomp}%
\usepackage{manyfoot}%
\usepackage{booktabs}%
\usepackage{algorithm}%
\usepackage{algorithmicx}%
\usepackage{algpseudocode}%
\usepackage{listings}%
\usepackage{adjustbox}
\usepackage{mathdots}

\theoremstyle{thmstyleone}%
\theoremstyle{thmstyletwo}%

\theoremstyle{thmstylethree}%

\begin{document}

\title[Article Title]{
  \begin{center}
    \textbf{ Pulse-driven photonic transitions and nonreciprocity in space-time modulated metasurfaces}
  \end{center}
}

\author[1,2]{\fnm{Zeki} \sur{Hayran}}\email{z.hayran@imperial.ac.uk}

\author[2]{\fnm{John B.} \sur{Pendry}}\email{j.pendry@imperial.ac.uk}

\author*[3]{\fnm{Prasad P.} \sur{Iyer}}\email{ppadma@sandia.gov}

\author*[1]{\fnm{Francesco} \sur{Monticone}}\email{francesco.monticone@cornell.edu}

\affil[1]{School of Electrical and Computer Engineering, Cornell University, Ithaca, New York 14853, USA}
\affil[2]{The Blackett Laboratory, Department of Physics, Imperial College London, London SW7 2AZ, United Kingdom}
\affil[3]{Center for Integrated Nanotechnologies, Sandia National Labs, Albuquerque, NM, USA}

\abstract{Time-varying photonic systems open new possibilities for controlling light, enabling photonic time crystals, time reflection and refraction, frequency conversion, synthetic gauge fields, optical nonreciprocity, among others. These effects emerge from the dynamic modulation of optical properties, which can mediate photonic transitions between eigenstates of different frequencies and/or wavevectors. To achieve such transitions, conventional approaches rely on periodic modulation schemes that demand ultrafast modulation rates and continuous energy input, posing significant practical challenges at optical frequencies. Here, we demonstrate that periodic-modulation-driven photonic transitions within the radiation continuum can be effectively mimicked using a single-period ultrafast pulse modulation, eliminating the need for sustained continuous modulation. By leveraging dispersion engineering in metasurfaces to tailor the density of states in the radiation continuum, we achieve controlled frequency transitions and theoretically demonstrate strong nonreciprocity for free-space waves as a key application. Our findings may guide future experimental research on time-varying photonics using materials such as transparent conductive oxides and semiconductors, expanding the possibilities for ultrafast and reconfigurable optical technologies. More broadly, our work may establish a practical and energy-efficient framework for dynamic photonic systems, with potential applications ranging from spatio-temporal wavefront manipulation to photonic computing and ultrafast signal processing.
}


\maketitle

\section{Introduction}\label{sec1}

The advent of spatially structured artificial electromagnetic media has profoundly expanded the means of manipulating wave phenomena, from guiding and localizing light to tailoring its interactions with matter. A pivotal milestone in this direction was Bykov’s 1975 work \cite{bykov1975spontaneous}, which demonstrated that spontaneous emission from an atom could be modified when embedded in a periodic dielectric structure featuring an electromagnetic band gap, an insight later extended into three dimensions by Yablonovitch \cite{yablonovitch1987inhibited} and John \cite{john1987strong}, leading to the development of the field of ``photonic crystals'' (PhCs) \cite{joannopoulos1997photonic}. These structures paved the way for unprecedented control over electromagnetic waves, alongside other key advances such as metamaterials \cite{pendry1998low, shelby2001experimental} and frequency-selective surfaces (FSSs) \cite{anwar2018frequency}. Collectively, these innovations---PhCs, metamaterials, and FSSs---have driven transformative advancements over recent decades \cite{zheludev2012metamaterials, simovski2020introduction, butt2021recent}, enabling precise control over wave propagation, dispersion, and localization by engineering the spatial dimensions of a system. As a further step in the quest to control light, in recent years the incorporation of time as a fourth dimension has been increasingly recognized as essential to fully harness the capabilities of structured media \cite{caloz2019spacetime, engheta2023four}, enabling dynamic control of electromagnetic properties beyond the limits of conventional time-invariant systems \cite{hayran2023using}. Yet, this extension into time-varying photonics presents new challenges, particularly at optical frequencies, where the practical implementation of strong and fast periodic temporal modulation is inherently demanding. Achieving nontrivial temporal periodicity effects often requires modulation speeds exceeding the oscillation period of the probe wave \cite{asgari2024photonic}. At such timescales, the most practical approach is to exploit optical nonlinearities to induce rapid refractive index changes \cite{bej2025ultrafast}. However, while nonlinear effects can, in principle, enable ultrafast modulation, the relatively weak nonlinear response of most optical materials \cite{soljavcic2004enhancement} typically results in a limited modulation strength at these high speeds, posing a fundamental challenge to practical implementations.

\begin{figure}
 \centering
 \begin{adjustbox}{width=0.78\linewidth, center}
  \includegraphics[keepaspectratio]{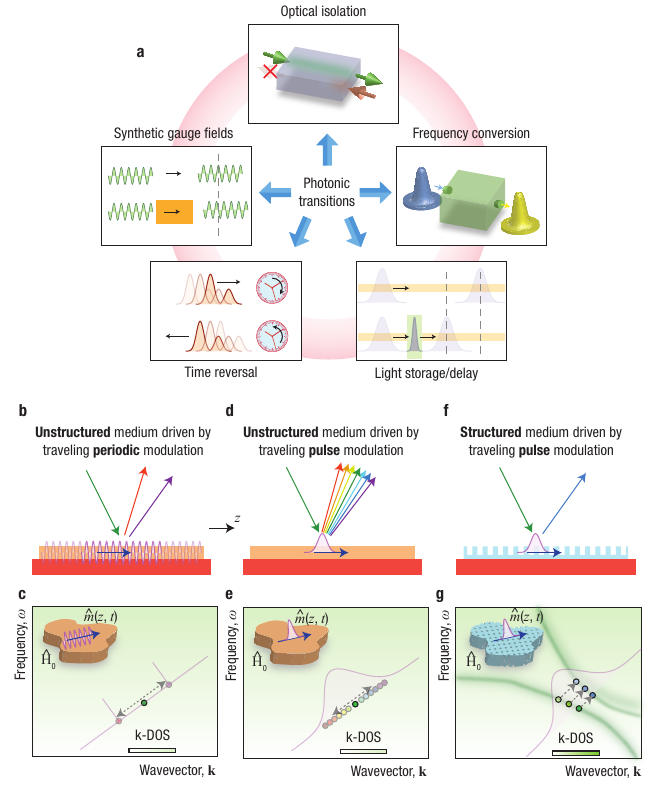}
  \end{adjustbox}
    \caption{\textbf{| Controlling photonic transitions under traveling pulse modulation.}  
\textbf{a,} Photonic transitions can be direct (frequency-only) or indirect (involving both frequency and wavevector changes), enabling functionalities such as isolation, frequency conversion, time reversal, light storage, and synthetic gauge fields.  
\textbf{b,c,} In a homogeneous medium, a traveling periodic modulation couples an input wave at a given \((\omega,k)\) to discrete radiation states. This selectivity comes from the fixed modulation frequency and wavevector.  
\textbf{d,e,} In contrast, a traveling pulse modulation has a broadband spectrum, so the transition spans many \((\omega,k)\) states. Without guided leaky modes (that is, without spectral regions of high DOS), the incident energy is spread inefficiently across the radiation continuum.
\textbf{f,g,} When the medium is structured to support guided leaky modes, the associated spectral regions of high DOS act as channels that the broadband pulse modulation can couple into. This allows efficient transitions without requiring precise tuning of the modulation frequency and wavevector. Unlike periodic modulation, which enforces discrete shifts, pulse modulation can couple into leaky modes even when the dispersions of the initial and final states are not parallel, since different frequency--wavevector offsets are allowed.}
    \label{fig1}
\end{figure}

To address these challenges, recent advances in time-varying photonic materials, particularly transparent conductive oxides such as indium tin oxide (ITO) \cite{zhou2020broadband} and aluminum zinc oxide (AZO) \cite{khurgin2020adiabatic}, as well as semiconductors like gallium arsenide (GaAs) \cite{karl2020frequency}, have opened new avenues for dynamic photonic systems. These materials exhibit ultrafast refractive index dynamics under strong optical excitation, with characteristic rise and decay times governed by their carrier dynamics. In general, they feature extremely fast rise times, often limited by the duration of the excitation pulse, enabling near-instantaneous modulation \cite{tirole2023double}. Their relaxation time is determined by underlying physical mechanisms such as free-carrier recombination or phonon interactions, which dictate how quickly the refractive index returns to its equilibrium state. Early studies reported relaxation times of several picoseconds \cite{reshef2019nonlinear}, but more recent work has demonstrated sub-picosecond recovery \cite{tirole2023double} and even modulation cycles with durations below 100 fs \cite{lustig2023time}, expanding their potential for high-speed optical applications. Despite these advances, such modulations remain limited to single \cite{zhou2020broadband, khurgin2020adiabatic} or few \cite{tirole2023double, harwood2025space} modulation periods, primarily due to energy constraints \cite{hayran2022homega}, making it impractical to achieve continuously modulated periodic time-variations with large refractive index changes in these material platforms. This is a considerable drawback, since periodic temporal modulation remains one of the most effective approaches for controlling light by inducing photonic transitions (see Fig. \ref{fig1}(a) for various applications of this effect), as it provides a well-defined modulation frequency and wave-vector, ensuring selective coupling between specific states as shown in Fig. \ref{fig1}(b,c) \cite{shaltout2015time, guo2019nonreciprocal, cardin2020surface, sisler2024electrically}. In contrast, when modulation is restricted to a single or few periods, the lack of well-defined spectral and momentum characteristics of the time-modulated medium leads to dispersion of an incident free-space wave across a broad continuum of frequencies and wave-vectors \cite{harwood2025space} (see Fig. \ref{fig1}(d,e)). Thus, while periodic time-modulation offers a promising route for wave control, its practical implementation remains challenging at optical frequencies, necessitating alternative approaches to achieve controlled photonic transitions between well-defined states under the above-mentioned constraints.

A compelling goal in time-varying photonics is therefore to replicate the effects of a periodic (spatio-)temporal modulation using a single or limited number of modulation cycles. To address this challenge, we drew inspiration from the well-established role of spatial structuring in controlling light emission, from tailoring spontaneous emission in photonic crystals to manipulating radiation in metamaterials and frequency-selective surfaces as mentioned above. By shaping the optical density of states (DOS), these structured media enable precise control over wave propagation, localization, and emission directionality. Driven by this, we explore whether similar structuring can be leveraged in time-varying photonics, where temporal modulation effectively acts as a new emission source, generating waves at different frequencies and wavevectors. We demonstrate that by engineering the DOS of a (spatio-)temporally modulated medium, one can induce controlled photonic transitions between well-defined modes, even with a single-period traveling pulse modulation. This control is achieved through spatial structuring, which shapes the available DOS and constrains transitions to specific spectral and angular regions (see Fig. \ref{fig1}(f,g)). We specifically apply this approach to modes within the radiation continuum, which remain accessible to free-space waves. Unlike most previous works in time-varying photonics, which have largely focused on (spatio-)temporally modulating homogeneous bulk media—often termed ``(space-)time metamaterials'' \cite{caloz2019spacetime, galiffi2022photonics, pendry2024qed}, where temporal modulation itself defines the metamaterial properties, our approach applies (spatio-)temporal modulation to pre-structured metamaterials or photonic crystals.

We stress that the application of temporal modulation to spatially structured media has a long history, from early demonstrations in fiber Bragg gratings and photonic crystals \cite{de1992optical, broderick1997optical, winn1999interband} to more recent investigations of temporally modulated periodic systems \cite{sivan2011theory, sivan2011time, karl2020frequency, wang2024expanding, hayran2025resonant}. These works laid the foundation for dynamic wave control in structured platforms, but their focus was largely on periodic modulation or guided-mode transitions. In parallel, spatiotemporal modulation has been examined in homogeneous dispersive media, including periodic modulation treated in the dispersive regime \cite{dana2014spatiotemporal} and pulse perturbations captured within a coupled-mode framework \cite{sivan2016coupled}. However, to the best of our knowledge, the regime of spatiotemporal pulse modulation in structured systems, particularly when targeting radiation continuum modes, has not been thoroughly explored. In this work we address this gap, showing that dispersion engineering of structured media allows a single-period pulse spatiotemporal modulation to mediate controlled and selective photonic transitions, an effect that has so far been associated mainly with continuous periodic modulation.

\section{Main}\label{sec2}

We now introduce a theoretical framework to describe the evolution of photonic states under spatiotemporal modulation. We represent the optical field by a state vector \(\lvert a(z,t)\rangle\), which collects the amplitudes of the unperturbed eigenmodes of the structure. This state can represent a superposition of both guided and radiation modes, and the bra--ket notation provides a compact way to describe its evolution. In the absence of modulation, the evolution of each component is governed by the unperturbed Hamiltonian \(\hat H_0\). When a traveling spatiotemporal modulation is applied, meaning a perturbation of the material parameters that varies in both space and time and whose profile moves through the structure with velocity \(v_{\text{m}}\), we describe its effect through an operator \(\hat m(t-z/v_{\text{m}})\). This operator captures the action of the modulation on the photonic state, allowing energy to be redistributed among different eigenmodes or among distinct frequency--wavevector components of the same mode. The combined effect of the unperturbed Hamiltonian and the traveling perturbation can then be written as

\begin{equation}
    i \left( \frac{\partial}{\partial z} + \frac{1}{v_{\text{m}}} \frac{\partial}{\partial t} \right) \lvert a(z,t) \rangle = \left[ \hat{H}_0 + \hat{m}(t - z/v_{\text{m}}) \right] \lvert a(z,t) \rangle .
    \label{eq1}
\end{equation}  

\noindent This formulation applies equally to discrete spectra and continua, since the eigenbasis of \(\hat H_0\) includes both guided branches and radiation states. We take \(\hat H_0\) and \(\hat m\) with units of inverse length, so that their eigenvalues match the longitudinal phase factor \(k-\omega/v_g\) of a plane-wave component, and normalize the state as \(\langle a(z,t)\lvert a(z,t)\rangle = 1\). Equation \ref{eq1} forms the starting point for our analysis. By projecting it onto the eigenbasis of \(\hat H_0\), one obtains a coupled--mode description in which the coefficients of the modes evolve according to the matrix elements of the modulation. This provides the foundation for analyzing transition probabilities in the following sections.

To simplify the analysis, we switch to a co-moving frame by defining new coordinates \( \zeta = z \) and \( \tau = t - z / v_{\text{m}} \), such that the evolution equation becomes

\begin{equation}
    i \, \partial_\zeta \lvert a(\zeta, \tau) \rangle = \left[ \hat{H}_0 + \hat{m}(\tau) \right] \lvert a(\zeta, \tau) \rangle.
    \label{eq2}
\end{equation}

\noindent Furthermore, we transition to the interaction picture, allowing the state \( \lvert a(\zeta, \tau) \rangle \) to be expressed as  
\begin{equation}
    \lvert a(\zeta, \tau) \rangle = e^{-i \hat{H}_0 \zeta} \lvert A(\zeta, \tau) \rangle.
    \label{eq3}
\end{equation}  

\noindent Substituting this into the time evolution equation transforms it into  

\begin{equation}
    i \partial_\zeta \lvert A(\zeta, \tau) \rangle = \hat{M}(\zeta, \tau) \lvert A(\zeta, \tau) \rangle,
    \label{eq4}
\end{equation}  

\noindent where \( \hat{M}(\zeta, \tau) = e^{i \hat{H}_0 \zeta} \hat{m}(\tau) e^{-i \hat{H}_0 \zeta} \) represents the modulation operator in the interaction picture. The state \( \lvert A(\zeta, \tau) \rangle \) can then be expressed as  

\begin{equation}
    \lvert A(\zeta, \tau) \rangle = \lvert A(\zeta_0, \tau) \rangle + \frac{1}{i} \int_{\zeta_0}^{\zeta} d\zeta' \, \hat{M}(\zeta', \tau) \lvert A(\zeta', \tau) \rangle.
    \label{eq5}
\end{equation}

\noindent Iterating this equation yields a Dyson-like series, which can be compactly expressed as  

\begin{equation}
    \lvert A(\zeta, \tau) \rangle = \mathcal{P} \exp \left[ -i \int_{\zeta_0}^{\zeta} d\zeta' \, \hat{M}(\zeta', \tau) \right] \lvert A(\zeta_0, \tau) \rangle.
    \label{eq6}
\end{equation}

Equation \ref{eq6} provides a formally exact solution to Eq. \ref{eq2}, describing the evolution of the photonic state under a \(\zeta\)-dependent interaction \( \hat{M}(\zeta, \tau) \). The path-ordering operator \( \mathcal{P} \) accounts for the non-commutativity of the interaction at different positions along \(\zeta\), ensuring that the evolution maintains the correct causal structure along the \(\zeta\)-direction.

To calculate the transition rate (i.e., the transmitted intensity into the final state per unit interaction length) between discrete photonic modes mediated by a traveling modulation, we evaluate the first-order contribution to the evolution operator in the interaction picture. We consider an initial state \( \lvert i \rangle = \lvert \Omega_1, \beta_1 \rangle \) and a final state \( \lvert f \rangle = \lvert \Omega_2, \beta_2 \rangle \), which form part of an orthonormal eigenbasis of the unperturbed Hamiltonian. Here, \( \Omega = \omega \) and \( \beta = k - \omega/v_{\text{m}} \) are the Fourier conjugates of \( \tau \) and \( \zeta \), respectively, with \( \omega \) and \( k \) corresponding to temporal and spatial frequencies in the laboratory frame.

Projecting the first-order term in Eq. \ref{eq6} onto the final state yields the transition amplitude to the final state as

\begin{equation}
    \alpha_{fi}^{(1)} = -i \int_{\zeta_0}^{\zeta} d\zeta' \, \langle f \lvert \hat{M}(\zeta', \tau) \rvert i \rangle.
    \label{eq7}
\end{equation}

\noindent We decompose the modulation operator as \( \hat{m}(\tau) = \hat{m}_0 \, V(\tau) \), where \( \hat{m}_0 \) encodes the spatial overlap between the eigenmodes of the unperturbed system and the modulation profile, thereby capturing the finite spatial extent and symmetry of the perturbation. The factor \( V(\tau) \) then specifies the temporal waveform of the modulation. The transition intensity is given by the squared magnitude of the amplitude

\begin{equation}
    |\alpha_{fi}^{(1)}|^2 = |\kappa_{fi}|^2\, |\langle f \lvert V(\tau) \rvert i \rangle|^2 \left|\int_{\zeta_0}^{\zeta} d\zeta'\, e^{i(\beta_f-\beta_i)\zeta'}\right|^2 ,
    \label{eq8}
\end{equation}

\noindent with the spatial coupling coefficient defined as \( \kappa_{fi} = \langle f \lvert \hat{m}_0 \rvert i \rangle \). For a long interaction region, the squared integral can be approximated as

\begin{equation}
    \left|\int_{\zeta_0}^{\zeta} d\zeta'\, e^{i(\beta_f-\beta_i)\zeta'}\right|^2 \approx (2\pi)\, (\zeta - \zeta_0) \, \delta(\beta_f-\beta_i).
    \label{eq9}
\end{equation}

\noindent Thus, the transition rate becomes

\begin{equation}
    \Gamma_{fi} = \frac{d}{d\zeta}\,|\alpha_{fi}^{(1)}|^2 \approx (2\pi)\, |\kappa_{fi}|^2\, |\tilde{V}(\Omega_f-\Omega_i)|^2\, \delta(\beta_f-\beta_i).
    \label{eq10}
\end{equation}

Equation \ref{eq10} serves as a classical analogue of Fermi’s golden rule, governing photonic state transitions driven by traveling space–time modulations. It reveals two key results: (i) the quantity \( \beta = k - \omega/v_{\text{m}} \) is conserved under a traveling modulation of the form \( \hat{m}(t - z/v_{\text{m}}) \) (in agreement with the recent Noether's theorem-based analysis in Ref. \cite{liberal2024spatiotemporal}), and (ii) to first order, transitions between frequency-separated states are determined by the spectral component \( \tilde{V}(\Omega_f - \Omega_i) \) of the modulation, where \( \Omega_i = \omega_i \) and \( \Omega_f = \omega_f \) are the frequencies of the initial and final states, respectively.

We next apply the general formalism to the case where the initial and final states belong to the radiation continuum. In this case, it is necessary to account for the density of available modes when constructing a wave packet. For a traveling modulation of the form \(\hat m(t-z/v_{\text{m}})\), the conserved quantity is \(\beta = k - \omega/v_{\text{m}}\). It is therefore natural to describe the continuum in terms of the variables \((\Omega,\beta)\), where \(\Omega=\omega\). We define \(\rho(\Omega,\beta)\) as the density of states at fixed \(\beta\), that is, the number of continuum modes per unit frequency along a constant-\(\beta\) slice. This is equivalent to the usual wavevector-resolved DOS, expressed in the mixed variables \((\Omega,\beta)\).  

With this definition, the excitation can be written as a wave packet over a finite continuum bandwidth \(\Delta\Omega_i\) at fixed \(\beta\). Each mode is weighted by the spectral amplitude \(A_i(\Omega)\), which specifies how the states are populated, and by the factor \(\sqrt{\rho(\Omega,\beta)}\), which accounts for the density of states

\begin{equation}
    \lvert A(\zeta_0, \tau) \rangle = \int_{\Delta\Omega_i} d\Omega \, \sqrt{\rho(\Omega,\beta)} \, A_i(\Omega) \, e^{-i\Omega \tau} \lvert \Omega,\beta \rangle,
    \label{eq11}
\end{equation}

\noindent with the normalization condition

\begin{equation}
    \int_{\Delta\Omega_i} \lvert A_i(\Omega) \rvert^2 \, \rho(\Omega,\beta) \, d\Omega = 1.
    \label{eq12}
\end{equation}

\noindent Since \(\beta\) is conserved under the traveling modulation, it is convenient to represent the continuum in terms of constant-\(\beta\) slices. This choice does not impose conservation but makes it explicit in the formulation, and it allows the analysis to focus directly on the subset of modes that are coupled by the modulation. Following the same reasoning as in the discrete case, in the long-\(\zeta\) interaction limit the total transition rate into the final continuum \(\Delta\Omega_f\) is given by

\begin{equation}
    \Gamma = (2\pi) \int_{\Delta\Omega_f} d\Omega_f \, \rho(\Omega_f,\beta) \, \left| \int_{\Delta\Omega_i} d\Omega_i \, \kappa(\Omega_i,\Omega_f,\beta) \, \sqrt{\rho(\Omega_i,\beta)} \, A_i(\Omega_i) \, \tilde{V}(\Omega_f-\Omega_i) \right|^2.
    \label{eq13}
\end{equation}

In Eq. \ref{eq13}, the inner integral represents a coherent sum over the initial continuum states, each contributing with an amplitude set by the spectral weight \(A_i(\Omega)\), the density-of-states factor \(\sqrt{\rho(\Omega,\beta)}\), and the coupling element \(\kappa(\Omega_i,\Omega_f,\beta)\) modulated by \(\tilde{V}(\Omega_f-\Omega_i)\). These contributions correspond to the same final mode and are summed coherently before the modulus is taken, while the outer integral accounts for distinct final states weighted by \(\rho(\Omega_f,\beta)\). The result is a classical analogue of Fermi’s golden rule for continuum states, which in the long-\(\zeta\) limit yields sharp and selective transitions in frequency and momentum.

It is instructive to contrast this situation with pulse modulation in homogeneous thin films \cite{jaffray2025spatio}. In that case, the short thickness of the film restricts the interaction to only a fraction of the modulation profile. The process is effectively an adiabatic time refraction, producing coupling into a few discrete angular and spectral channels, but with weak conversion because of the limited interaction time. In structured systems, by contrast, the modulation can interact with the optical field over longer distances, enabling stronger conversion and redistribution across a broader set of spatio-spectral states.

For clarity, throughout this work we use the term \textit{traveling pulse modulation} to describe a temporally localized modulation profile that propagates across the structure, spans approximately one modulation cycle, and has a broadband spectrum in frequency and wavevector. The alternative terms \textit{broadband modulation} and \textit{single-period ultrafast pulse modulation} are used interchangeably. Here, ``ultrafast'' refers to modulation dynamics relative to the optical cycle of the probe wave: the rise time can be shorter than a single optical period at near-infrared or visible frequencies (tens of femtoseconds), while the decay can extend to hundreds of femtoseconds or a few picoseconds but still remains in the ultrafast regime. This definition concerns the temporal duration of the modulation, not the velocity at which it propagates.

\subsection{Special Case: Periodic Modulation}

Equation \ref{eq13} can be greatly simplified when the modulation is \(\tau\)-periodic, specifically, when it is sinusoidal with a modulation frequency \(\Omega_{\mathrm{mod}} = \omega_{\mathrm{mod}}\). In this case, the Fourier transform of the modulation is nonzero only at the discrete frequencies \(\pm \Omega_{\mathrm{mod}}\). Accordingly, we can write

\begin{equation}
    \tilde{V}(\Omega_f-\Omega_i) = \tilde{V}(\pm \Omega_{\mathrm{mod}}) \, \delta(\Omega_f-\Omega_i \mp \omega_{\mathrm{mod}}).
\end{equation}

\noindent Substituting this expression into Eq. \ref{eq13} yields the differential transition rate

\begin{equation}
    \frac{d\Gamma}{d\omega_i\,d\omega_f} = (2\pi) \, \rho(\omega_i,\beta) \, \rho(\omega_f,\beta) \, \left| \kappa(\omega_i,\omega_f,\beta) \, A_i(\omega_i) \, \tilde{V}(\pm \omega_{\mathrm{mod}}) \right|^2 \, \delta(\omega_f-\omega_i \mp \omega_{\mathrm{mod}}).
\end{equation}

\noindent This result rigorously demonstrates that each input spectral component \(A_i(\omega_i)\) couples exclusively to a final state at frequency \(\omega_f = \omega_i \pm \omega_{\mathrm{mod}}\). In other words, the periodic modulation enforces a strict frequency conversion process by permitting only a discrete frequency shift of \(\pm \omega_{\mathrm{mod}}\) in the transition process, with a corresponding wavevector change of \(\Delta k = \pm \omega_{\mathrm{mod}}/v_{\text{m}}\) to conserve the quantity \(\beta = k - \omega/v_{\text{m}}\) (see Fig. \ref{fig1}(b,c)).

\subsection{Special Case: Narrow Density of States}

We now consider the regime where the density of states in both the initial and final continua exhibits narrow maxima, while the modulation spectrum remains broadband. In this situation, the DOS varies appreciably only within a small frequency window around a maximum \(\omega_{0,i}\) or \(\omega_{0,f}\), and the input spectral amplitude is also confined within this interval. Under these conditions, in Eq. \ref{eq13} the DOS and input amplitude can be approximated by their values at the peak frequencies, leading to the simplified transition rate  

\begin{equation}
    \Gamma \approx (2\pi)\, \rho(\omega_{0,i},\beta) \rho(\omega_{0,f},\beta) \, \left| \kappa(\omega_{0,i},\omega_{0,f},\beta) \, A_i(\omega_{0,i}) \, \tilde{V}(\omega_{0,f}-\omega_{0,i}) \right|^2.
    \label{eq16}
\end{equation}

This expression shows explicitly how the transition rate is weighted by the DOS at the participating frequencies and by the spatial coupling coefficient \(\kappa(\omega_{0,i},\omega_{0,f},\beta)\). In practice, large values of DOS are often associated with resonant or leaky modes that confine energy inside the structure, which in turn increases the overlap with the modulation and enhances \(\kappa\). The dominant transition channels are therefore set jointly by the DOS and by the coupling coefficient, rather than by the modulation spectrum alone.

In this regime, the frequency selectivity resembles that of periodic modulation, where a single modulation frequency enforces a discrete frequency shift. The crucial difference is that periodic modulation requires careful tuning of both the modulation frequency and wavevector to achieve phase matching with a specific transition. In contrast, a traveling pulse modulation provides a range of Fourier components, so the interaction can automatically couple into whichever states satisfy the conservation laws, without external tuning, provided that the required components are present in the modulation spectrum. The efficiency of these transitions is then governed by the structural properties of the system, which set both the density of states and the coupling strength. Hence, unlike approaches based on monochromatic traveling-wave modulations \cite{winn1999interband, yu2009complete, fang2012photonic, shi2017optical}, traveling pulse modulation enables multiple frequency shifts within the same interaction, acting as if several modulation frequencies and wavevectors were applied at once.

Finally, although Eq. \ref{eq16} was derived from first-order perturbation theory, the sharp-DOS regime ensures that the main conclusions hold beyond first order. Because both the input and output continua are confined to narrow spectral regions, the dominant contribution always comes from modulation components matching the frequency difference between the states. Higher-order terms therefore reinforce the same selection rule, and the transition rate remains governed by the DOS at the final frequency, preserving the Fermi’s golden rule--like behavior.

\begin{figure}
 \centering
 \begin{adjustbox}{width=1.2\linewidth, center}
  \includegraphics[keepaspectratio]{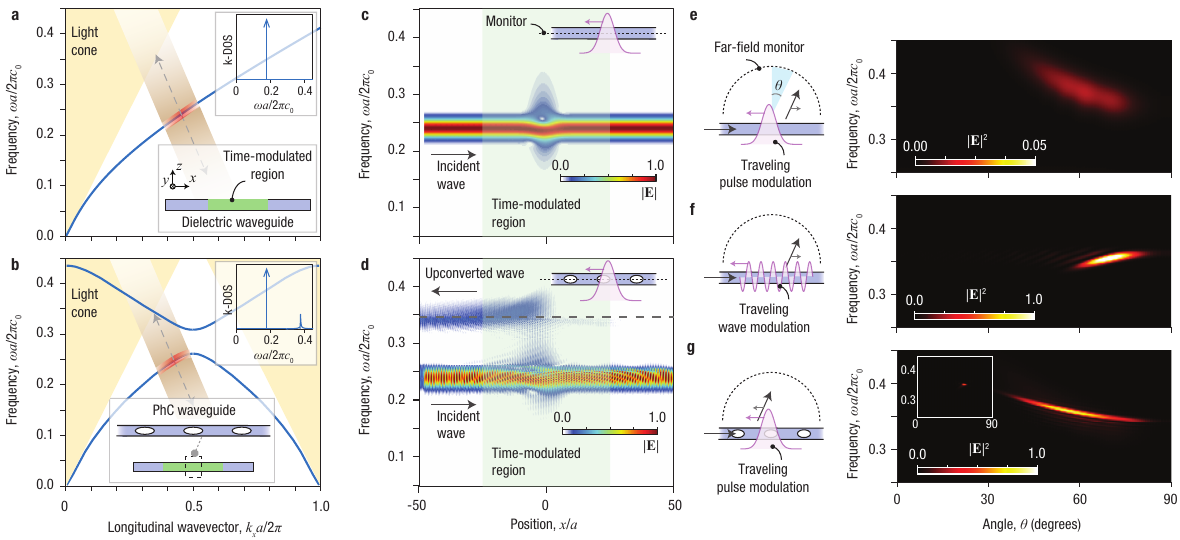}
  \end{adjustbox}
    \caption{\textbf{| Photonic transitions in slab and photonic crystal waveguides under traveling pulse modulation.}  
\textbf{a,} Dispersion diagram of a dielectric slab waveguide showing the fundamental guided mode and the diagonal transition induced by a traveling pulse modulation.
\textbf{b,} Dispersion of a PhC slab waveguide, obtained by introducing periodic holes, showing band folding and the appearance of a leaky mode inside the light cone (yellow region). The upper-right insets in \textbf{a} and \textbf{b} show the wavevector-resolved DOS at \(k_x a / 2\pi = 0.25\); the lower insets illustrate the corresponding structures.
\textbf{c,} Frequency-resolved electric field magnitude along the waveguide axis for the homogeneous slab, showing mostly spectral broadening and weak radiation since no true leaky modes are available.  
\textbf{d,} For the PhC waveguide, the traveling pulse modulation couples the guided mode to the leaky mode inside the light cone (red circle), producing efficient radiation. The leaky mode propagates in the negative \(x\)-direction, consistent with the negative slope of its dispersion relation.
\textbf{e--g,} Far-field intensity distributions for different cases.
\textbf{e,} Homogeneous slab with traveling pulse modulation: broad and inefficient radiation over many angles and frequencies, reflecting the continuous radiation continuum.  
\textbf{f,} Homogeneous slab with periodic traveling-wave modulation: discrete coupling into specific radiation modes at selected \((\omega,k)\) points.
\textbf{g,} PhC slab with traveling pulse modulation: efficient coupling into specific frequencies and angles set by the leaky mode dispersion. The far-field amplitude is more than an order of magnitude stronger than in \textbf{e} and comparable to \textbf{f}. The inset shows the case of periodic modulation combined with spatial structuring, where coupling is restricted to a single frequency--angle channel due to the monochromatic modulation spectrum.}
    \label{fig2}
\end{figure}

\begin{figure}
 \centering
 \begin{adjustbox}{width=1.2\linewidth, center}
  \includegraphics[keepaspectratio]{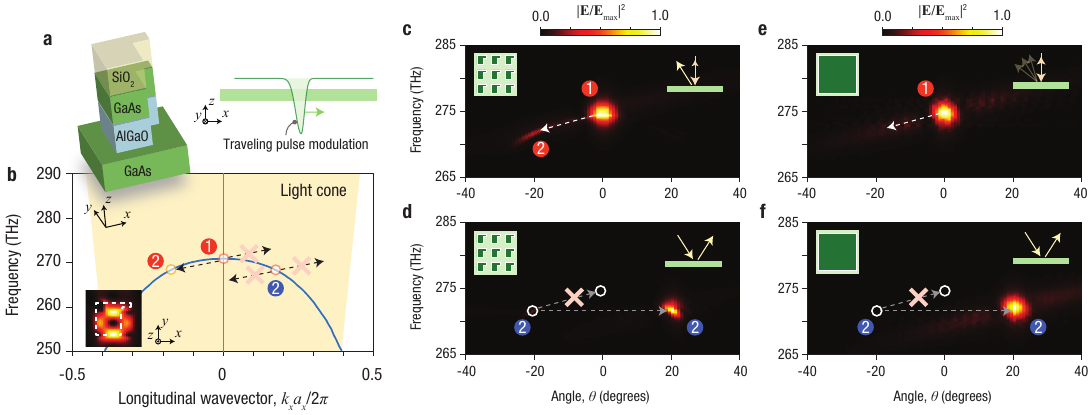}
  \end{adjustbox}
    \caption{\textbf{| Metasurface-mediated nonreciprocal photonic transitions under a traveling pulse modulation with fast relaxation.}  
\textbf{a,} Unit cell of the metasurface comprising GaAs, AlGaO, and SiO$_2$, with a traveling pulse modulation applied along the \(+x\)-direction.
\textbf{b,} Dispersion diagram of the metasurface, showing transitions (dashed arrows) between states 1 and 2 induced by the modulation. The inset shows the magnetic field profile of state 1.
\textbf{c,} Far-field intensity for the metasurface, demonstrating excitation of state 1 (red circle) and its transition to state 2 (red circle), radiating at a different angle and frequency.
\textbf{d,} In the backward direction, excitation of state 2 does not couple back to state 1. Because the modulation enforces diagonal transitions in \((\omega, k_x)\) space, the backward excitation lies on the opposite side of the dispersion (positive \(k_x\)), and the same diagonal shift cannot reach state 1. As a result, mainly specular reflection occurs (blue circle).
\textbf{e,} For a homogeneous slab without metasurface structuring, state 1 couples very weakly to state 2, since the frequency--angle points of state 1 and state 2 do not correspond to any leaky guided mode in the radiation continuum.
\textbf{f,} In the homogeneous case, backward excitation at the frequency--angle of state 2 also results mainly in specular reflection, confirming that metasurface structuring is required for strong nonreciprocal transitions.}
    \label{fig3}
\end{figure}

\section{Results} \label{sec3}

\section*{Results}

To validate the above insights, we examined representative structures that illustrate how modifying the DOS through spatial structuring enables selective photonic transitions under traveling pulse modulation. Unlike periodic modulation, which has a well-defined modulation frequency and wavevector, a traveling pulse carries a broad spectrum in both frequency and wavevector. Our aim is to show how dispersion engineering can channel this broadband drive into targeted transitions. As a reference, we first analyze a conventional dielectric slab waveguide, followed by the introduction of spatial structuring in the form of a photonic crystal slab to assess its impact on transition dynamics.

Figure \ref{fig2}(a) shows the dispersion diagram for the fundamental transverse electric (TE) mode of a conventional two-dimensional dielectric slab waveguide (see Methods section for structural details). A traveling pulse modulation was applied to a specific region of the waveguide, and the probe pulse was kept within this region during the interaction. The modulation propagates opposite to the waveguide mode, inducing diagonal transitions in \((\omega,k)\) space. These transitions proceed toward negative \(k\) and positive \(\omega\), or vice versa, as indicated by the dashed arrow in Fig. \ref{fig2}(a), enabling coupling between the guided mode and the radiation continuum (yellow area). Because the modulation spectrum is broadband in both frequency and wavevector, it couples into all accessible states of the continuum. Since the continuum spans a wide range of \(\omega\) for each longitudinal component \(k_x\), this pulse-induced coupling leads to inefficient and non-selective transitions, as discussed earlier.

To address this, we introduced spatial structuring by transforming the dielectric slab waveguide into a photonic crystal (PhC) slab waveguide, as shown in Fig. \ref{fig2}(b) (see Methods section for structural details). This structuring produces a leaky waveguide mode that partially resides within the light cone, giving rise to a Lorentzian-like wavevector-resolved DOS due to radiation leakage \cite{ohtaka2004derivation} (see the insets of Fig. \ref{fig2}(a,b)). Because the dispersions of the guided and leaky modes are not parallel, there is no single modulation frequency and wavevector that can couple them efficiently. A traveling pulse modulation, being broadband in both frequency and wavevector, supplies the necessary spectral components without precise tuning and effectively ``scans'' the available phase space to excite the leaky mode inside the light cone.  

To validate this prediction, we conducted finite-difference time-domain (FDTD) simulations, with results shown in Figs. \ref{fig2}(c) and \ref{fig2}(d) for the homogeneous and PhC waveguide cases, respectively. Figure \ref{fig2}(c) displays the spectrally resolved electric field magnitude along the waveguide axis in the homogeneous case, showing no evidence of selective coupling. By contrast, Fig. \ref{fig2}(d) demonstrates selective excitation of the leaky mode in the PhC waveguide. The dashed line in Fig. \ref{fig2}(d) marks the portion of the leaky mode that lies inside the light cone and therefore couples to radiation. Notably, the upconverted wave exhibits a negative group velocity relative to the input waveguide mode, consistent with the negative slope of the leaky-mode dispersion, while radiating with a positive longitudinal wavevector \(k_x\), since only the portion of the mode inside the light cone contributes to radiation.

Figures \ref{fig2}(e)--\ref{fig2}(g) compare the far-field intensity profiles for different modulation and waveguide configurations, highlighting the role of spatial structuring. In a homogeneous slab waveguide under traveling pulse modulation (Fig. \ref{fig2}(e)), the radiation spreads across a broad spectral and angular range, indicating weak and non-selective coupling. Introducing periodic modulation in the same homogeneous waveguide (Fig. \ref{fig2}(f)) produces selective coupling into specific radiation channels, but at the cost of continuous modulation and precise tuning. The PhC slab waveguide under traveling pulse modulation (Fig. \ref{fig2}(g)) achieves comparable selectivity without these constraints. Here the broadband modulation spectrum can couple states with different frequency--wavevector offsets, so the dispersions of the initial and final states do not need to be parallel. The PhC dispersion then shapes which of these offsets are allowed, thereby controlling the spectral and angular distribution of the emission. For reference, the inset of Fig. \ref{fig2}(g) shows periodic modulation combined with spatial structuring, where the monochromatic drive restricts coupling to a single spectral component. Overall, this comparison demonstrates that a traveling pulse modulation, when combined with dispersion engineering, can reproduce and extend the selective functionality of a periodic modulation without requiring tuning or sustained energy input.

\begin{figure}
 \centering
 \begin{adjustbox}{width=1.2\linewidth, center}
  \includegraphics[keepaspectratio]{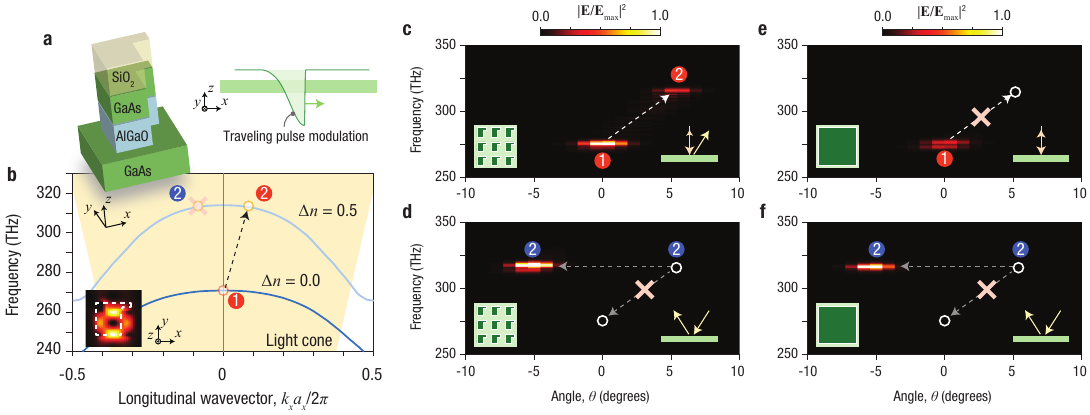}
  \end{adjustbox}
    \caption{\textbf{| Metasurface-mediated nonreciprocal photonic transitions under a traveling pulse modulation with a slow relaxation.} 
\textbf{a,} Unit cell of the metasurface comprising GaAs, AlGaO, and SiO$_2$, with a traveling pulse modulation applied along the \(+x\)-direction. 
\textbf{b,} Dispersion diagram of the metasurface for two GaAs refractive index states (\(\Delta n = 0\) and \(\Delta n = 0.5\)), showing how the slow relaxation produces an effective transition between them. The lower left inset shows the magnetic field profile of state 1. 
\textbf{c,} Far-field intensity for the metasurface, showing excitation of state 1 (red circle) and its coupling to state 2 (red circle) enabled by the modulation. 
\textbf{d,} In the backward direction, the excitation begins from the unpumped configuration (\(\Delta n = 0\)), where the frequency–angle point of state 2 does not lie on any dispersion branch. No coupling back to state 1 occurs and only specular reflection is observed (blue circle). 
\textbf{e,} For a homogeneous slab without metasurface structuring, state 1 does not couple to state 2, since the frequency–angle point of state 2 does not correspond to any leaky guided mode in the radiation continuum. 
\textbf{f,} In the homogeneous case, backward excitation at the frequency–angle of state 2 also produces only specular reflection, as no dispersion branch exists there, similar to \textbf{d}.
}
    \label{fig4}
\end{figure}

Having established how traveling pulse modulation combined with spatial structuring can control transitions in guided-wave systems, we now turn to open systems to explore nonreciprocal control of free-space radiation. Spatial structuring provides the dispersion features needed to define radiation states, while the traveling pulse modulation supplies a broadband drive that enables coupling between them. Building on previous work \cite{karl2020frequency}, we consider a metasurface platform where this mechanism produces nonreciprocal beam deflection together with frequency conversion, a functionality with strong practical applications in free-space optical isolation and directional signal processing \cite{caloz2018electromagnetic}.  

The metasurface unit cell is shown in Fig. \ref{fig3}(a), and its dispersion diagram near a quasi-bound state in the continuum \cite{koshelev2018asymmetric} is given in Fig. \ref{fig3}(b). Figures \ref{fig3}(c) and \ref{fig3}(d) show the simulated far-field intensity distributions as a function of angle and frequency. At normal incidence, excitation of a leaky mode (state 1, red circles in Fig. \ref{fig3}(b,c)) couples to the same leaky mode at a different point of the dispersion diagram (state 2, red circles in Fig. \ref{fig3}(b,c)), demonstrating a controlled transition within the radiation continuum. Each ``state'' refers to a wave packet localized in frequency and angle around a dispersion point. In the backward direction, excitation at the frequency--angle of state 2 (blue circles in Fig. \ref{fig3}(d)) does not couple back to state 1. Because the modulation enforces diagonal transitions in \((\omega,k_x)\) space, the backward excitation lies on the opposite side of the dispersion and cannot connect to state 1. As a result, only specular reflection occurs, yielding an isolation ratio of 25 dB at the central frequency and angle, and demonstrating strong optical nonreciprocity.

To highlight the role of spatial structuring, we compare the metasurface response with that of an equivalent homogeneous slab subject to the same traveling pulse modulation (see Methods section for structural details). In the homogeneous case, all guided modes lie outside the light cone, so there are no leaky resonances available for the modulation to couple into. As a result, state 1, representing the normally incident excitation, couples only very weakly to other radiation-continuum states, producing faint nonreciprocal spectral and angular scattering (see Fig. \ref{fig3}(e)). The response is therefore dominated by reflection. In the backward direction, excitation at the frequency--angle of state 2 also produces only specular reflection (see Fig. \ref{fig3}(f)). This confirms the absence of strong nonreciprocal behavior and shows that spatial structuring is the essential ingredient for enabling strong nonreciprocity under traveling pulse modulation, without requiring periodic temporal modulation.

So far we considered traveling pulse modulations with fast decay, where the material relaxes back to its original eigenstates while the wave is still inside the medium. We now study the opposite case, where the decay is slow. During the interaction window the system does not relax to its initial eigenstates and it resides in a transient configuration. This connects our analysis with front induced transitions \cite{gaafar2019front} and extends it to radiation modes. In this regime, Eq. \ref{eq16} does not apply, since the dominant transition is between the original state and the transiently excited state.

To illustrate this mechanism, we use the same metasurface as in Fig. \ref{fig3}, now subject to a slow-decay traveling pulse modulation (Fig. \ref{fig4}(a)). The corresponding dispersion is shown in Fig. \ref{fig4}(b). The slow decay creates two effective refractive index configurations: the unpumped state (\(\Delta n = 0\)) and the transient excited state (\(\Delta n = 0.5\)). The dynamics can therefore be approximated as coupling between these two states, which is the same physical principle underlying front-induced transitions, here applied to free-space radiation to achieve nonreciprocal beam deflection. To our knowledge, this represents the first demonstration of front-induced nonreciprocity for free-space waves.  

The simulated far-field intensity distributions are shown in Figs. \ref{fig4}(c) and \ref{fig4}(d). At normal incidence, excitation of state 1 (red circles in Fig. \ref{fig4}(b,c)) couples into state 2 of the transient configuration (red circles in Fig. \ref{fig4}(b,c)). In the backward direction, excitation at the frequency--angle of state 2 encounters the unpumped configuration, where the frequency--angle point of state 2 does not lie on any dispersion branch. As a result, no coupling back to state 1 occurs, and only specular reflection is observed (blue circles in Fig. \ref{fig4}(d)). This asymmetry produces strong nonreciprocity with an isolation ratio of 61 dB.

As a reference, we carried out the same analysis for an equivalent homogeneous slab (Figs. \ref{fig4}(e,f)) (see Methods section for structural details). In this case, state 1 remains largely uncoupled since no leaky modes are available, and the scattered radiation is very weak. In the backward direction, excitation at the frequency--angle of state 2 also results mainly in specular reflection. These results confirm that spatial structuring is essential for enabling strong nonreciprocal transitions under slow-decay traveling pulse modulation.

\section{Conclusion}\label{sec4}

In summary, we have shown that traveling, single-period ultrafast pulse modulations, when combined with spatial structuring, provide a powerful route to control photonic transitions. Unlike approaches based on continuous or periodic modulations, which face practical challenges in sustaining strong index changes over long durations at optical frequencies, a single ultrafast modulation can selectively couple guided and radiation states, control the spectral and angular properties of leaky-wave emission, and enable robust nonreciprocity. These capabilities arise from the interplay between the broadband modulation spectrum and dispersion-engineered DOS, which together define the accessible transition channels.

The findings presented here point to promising applications. Nonreciprocal beam steering and frequency conversion can benefit free-space optical isolation and directional signal processing, while selective coupling into leaky states suggests opportunities in light detection, sensing, and on-chip quantum photonics. At a more fundamental level, the results show how metasurfaces with momentum-dependent responses, commonly associated with nonlocal effects \cite{shastri2023nonlocal, monticone2025nonlocality}, can be combined with ultrafast time variations to realize wave phenomena that are inaccessible with time-invariant structures. Together with ongoing progress in ultrafast material platforms \cite{reshef2019nonlinear} and photonic time crystals \cite{wang2024expanding, hayran2025resonant}, these findings outline a pathway toward experimentally realizable, time-varying photonic systems that exploit single-period modulations for both practical functionalities and new regimes of wave physics.

\section*{Methods}

\bmhead{Spatio-temporal modulation design}

In the theory, the modulation enters as \(\hat m_0 V(\tau)\). The operator \(\hat m_0\) specifies how the perturbation acts on the spatial field distribution, including localization in space and the modal coupling it produces, and has units of inverse length. The factor \(V(\tau)\) is a dimensionless temporal waveform in the co moving variable \(\tau = t - z/v_m\). In the simulations, we implement the same modulation as a refractive index change \(\Delta n\, s(\mathbf r)\, V(x,t)\), where \(\Delta n\) is the modulation amplitude, \(s(\mathbf r)\) is a spatial mask that selects the modulated region, and \(V(x,t)\) is the same normalized waveform written in the lab frame as \(V(x,t) = V(t - x/v_m)\).

To model realistic material response in GaAs, ITO, and AZO, we use an asymmetric pulse with different rise and fall durations. The normalized waveform is

\begin{equation}
V(x,t) = 
\begin{cases}
\exp \left( -\dfrac{(t - x/v_{\mathrm{m}})^2}{2\sigma_{\mathrm{rise}}^2} \right), & (t - x/v_{\mathrm{m}}) \leq t_0 \\[10pt]
\exp \left( -\dfrac{(t - x/v_{\mathrm{m}})^2}{2\sigma_{\mathrm{fall}}^2} \right), & (t - x/v_{\mathrm{m}}) > t_0,
\end{cases}
\end{equation}

\noindent where \(v_{\mathrm m}\) is the modulation group velocity, \(\sigma_{\mathrm{rise}}\) and \(\sigma_{\mathrm{fall}}\) set the rise and fall durations, and \(t_0\) marks the temporal center. The full perturbation used in the simulations is \(\Delta n\, s(\mathbf r)\, V(x,t)\), which maps directly to the theoretical form \(\hat m_0 V(\tau)\).

\bmhead{Structural design for the leaky waveguide}

The dielectric slab waveguide studied in Fig. \ref{fig2} is a two-dimensional structure with a thickness of \(0.22a\) and a refractive index of 3.5. The region subjected to temporal modulation is limited to a length of \(25a\) along the waveguide axis, which in the notation of the previous section corresponds to the spatial mask \(s(\mathbf r)\). For the homogeneous slab, \(s(\mathbf r)\) equals unity throughout this region. For the PhC waveguide, the same mask is applied but with the elliptical holes excluded, so that only the GaAs material is modulated while the air holes remain unperturbed.  

The PhC waveguide is formed by introducing elliptical holes with a major radius of \(0.2a\) along the \(x\)-direction and a minor radius of \(0.075a\) along the \(z\)-direction, arranged periodically with a period of \(a\). The temporal modulation is implemented as a symmetric Gaussian traveling pulse, moving along the \(-x\)-direction with a group velocity of \(1.18c_0\). The pulse modulation has a temporal standard deviation of \(\sigma_\mathrm{rise} = \sigma_\mathrm{fall} = 0.72a/c_0\) and a modulation amplitude of \(\Delta n = 1\). 

The periodic spatiotemporal modulation is implemented as a sinusoidal variation of the refractive index, centered at \(n = 3.5\). The modulation amplitude is \(\Delta n = 1\) for the homogeneous slab and \(\Delta n = 0.1\) for the PhC waveguide. The modulation frequency is chosen as \(0.12c_0/a\), and the modulation wavevector as \(0.1 \, 2\pi/a\).

\bmhead{Structural design for optical nonreciprocity}

The structural details used in Fig. \ref{fig4} are the same as in Ref. \cite{karl2020frequency}. The temporal modulation is applied only to the GaAs unit cell and not to the GaAs substrate. This corresponds to choosing the spatial mask \(s(\mathbf r)\) as unity inside the GaAs unit cell and zero elsewhere, so that the refractive index perturbation is \(\Delta n\, s(\mathbf r)\, V(x,t)\). In Fig. \ref{fig3}, the modulation propagates along the \(x\)-direction with a group velocity of \(0.03c_0\) and exhibits a standard deviation of \(\sigma_\mathrm{rise} = 32\) fs for the excitation phase and \(\sigma_\mathrm{fall} = 55\) fs for the relaxation phase. In Fig. \ref{fig4}, the modulation travels along the \(x\)-direction with a group velocity of \(1.41c_0\), with the same excitation duration of \(\sigma_\mathrm{rise} = 32\) fs, while the relaxation phase is significantly longer, with \(\sigma_\mathrm{fall} = 1000\) fs. 

For the equivalent homogeneous slab used in Figs. \ref{fig3} and \ref{fig4}, all nanostructured layers shown in Figs. \ref{fig3}(a) and \ref{fig4}(a) are extended uniformly in the \(xy\)-plane so that the structure becomes spatially homogeneous while retaining the same vertical layer composition.

\bmhead{Simulation details}

All simulations were performed using Lumerical FDTD Solutions. The time-varying materials were implemented as material plugins utilizing the Flexible Material Plugin feature. In Figs. \ref{fig3} and \ref{fig4}, periodic boundary conditions were applied in the transverse $z$-direction. Additionally, for Fig. \ref{fig3} and \ref{fig4}, periodic boundary conditions were used in the transverse $y$-direction, with a finite number of 81 unit cells along the $x$-direction.

\backmatter

\bmhead{Acknowledgments}
This work was performed, in part, at the Center for Integrated Nanotechnologies, an Office of Science User Facility operated for the U.S. Department of Energy (DOE) Office of Science. Sandia National Laboratories is a multimission laboratory managed and operated by National Technology \& Engineering Solutions of Sandia, LLC, a wholly owned subsidiary of Honeywell International, Inc., for the U.S. DOE’s National Nuclear Security Administration under contract DE-NA-0003525. This paper describes objective technical results and analysis. Any subjective views or opinions that might be expressed in the paper do not necessarily represent the views of the U.S. Department of Energy or the United States Government.

\section*{Declarations}

\bmhead{Funding}
FM and ZH acknowledge support from the US Department of Energy (through Sandia National Laboratories) with grant no. DE-NA-0003525. ZH and JBP acknowledge support from the UKRI Engineering and Physical Sciences Research Council (EPSRC) (EP/Y015673/1).

\bmhead{Conflict of interest/Competing interests}
The authors declare no competing interests.

\bmhead{Availability of data and materials}
Authors can confirm that all relevant data are included in the article.

\bmhead{Code availability}
The code used to produce these results is available upon request from the corresponding author.

\bibliography{sn-bibliography}

\end{document}